\documentclass[a4paper,11pt]{article}
\usepackage{pos}

\renewcommand{\logo}{\relax}

\title{STeVECat, the Spectral TeV Extragalactic Catalog}

\author*[a]{Lucas Gréaux}
\author[a]{Jonathan Biteau}
\author[b]{Tarek Hassan}
\author[c]{Olivier Hervet}
\author[d,e]{Mireia Nievas Rosillo}
\author[c]{David A. Williams}

\affiliation[a]{Université Paris-Saclay, CNRS/IN2P3, IJCLab,\\
    91405 Orsay, France}
\affiliation[b]{Centro de Investigaciones Energéticas, Medioambientales y Tecnológicas, \\
    E-28040 Madrid, Spain}
\affiliation[c]{ Santa Cruz Institute for Particle Physics and Department of Physics, University of California, \\
    Santa Cruz, CA 95064, USA}
\affiliation[d]{Instituto de Astrofísica de Canarias,\\
    E-38205 La Laguna, Spain}
\affiliation[e]{Universidad de La Laguna, Dept. Astrofísica,\\
    E-38206 La Laguna, Tenerife, Spain}

\emailAdd{lucas.greaux@ijclab.in2p3.fr}
\emailAdd{jonathan.biteau@ijclab.in2p3.fr}
\emailAdd{tarek.hassan@ciemat.es}
\emailAdd{ohervet@ucsc.edu}
\emailAdd{mireia.nievas@iac.es}
\emailAdd{daw@ucsc.edu}

\newcommand{\comment}[1]{}

\abstract{
    The three main collaborations operating the current generation of imaging atmospheric Cherenkov telescopes (IACTs: H.E.S.S., MAGIC, VERITAS) publish their gamma-ray data in different formats and repositories. Extragalactic sources are highly variable at very-high energies (VHE, $E>100\,$GeV), and a unified repository would enable joint analyses of collections of extragalactic VHE spectra. To this aim, we have developed the Spectral TeV Extragalactic Catalog, STeVECat, which gathers high-level products of IACT observations from 1992 to 2021. We selected all publications in journals referenced in TeVCat that presented archival spectra with at least two points. We compiled the corresponding spectral data and formatted them following the convention adopted in available public repositories (GammaCat and VTSCat). In addition to spectral points with associated physical units, we provide meta-data featuring observation periods, livetime, excess counts over background and significance, as well as the coordinates, types and redshifts of the sources whenever available. STeVECat combines observations from 173 journal publications, compared to 72 in the previous reference compilation of extragalactic gamma-ray spectra (Biteau \& Williams, 2015). STeVECat is the most extensive set of VHE extragalactic spectra collected so far, with 403 spectra from 73 sources. The full catalog can readily be loaded with GammaPy, the Science Analysis Tool selected by the Cherenkov Telescope Array Observatory. Our compilation efforts enable population studies of extragalactic gamma-ray sources, studies of the GeV-TeV connection, and studies of absorption on the extragalactic background light.
}

\FullConference{%
  7th Heidelberg International Symposium on High-Energy Gamma-Ray Astronomy (Gamma2022)\\
  4-8 July 2022\\
  Barcelona, Spain\\}

\begin{document}
\maketitle

\section{Introduction}

The first detection of TeV photons from an extragalactic object, Markarian 421, by the Whipple Observatory in 1992 marked the beginning of extragalactic gamma-ray astronomy at very-high energy (VHE, $E>100$\,GeV) \cite{Punch_92}.
From there, numerous VHE extragalactic sources have been discovered by ground based imaging atmospheric Cherenkov telescopes (IACTs).  
Their sensitivity has greatly improved over the last two decades, and they were able to demonstrate that extragalactic sources are highly variable at VHE \cite{2022Galax..10...35P}.

Extragalactic sources detected at VHE are mostly active galactic nuclei (AGNs), whose emission spans wavelengths from radio to gamma rays.
The component generally assumed to come from inverse Compton radiation peaks in the gamma-ray band.
Contemporaneous observations at GeV and TeV are thus key to understand AGNs \cite{2022Galax..10...35P}.
There are nearly a hundred times more extragalactic sources detected in the GeV band than at TeV energies, making large population studies at VHE difficult without adequate spectral corpora \cite{Gamma-ray_cosmo}.
Moreover, the extragalactic VHE spectra observed from Earth are not the spectra emitted by the sources.
The extragalactic background light (EBL, see \cite{BW15}, hereafter BW15) induces absorption features on the observed spectra, which can be studied to probe the electromagnetic content of the Universe. 
Population studies of extragalactic VHE sources, studies of the GeV-TeV connection and EBL studies with TeV gamma-rays all require a large corpus of extragalactic observations \cite{Gamma-ray_cosmo, NievasRosillo:2021aS}.

The current generation of IACTs is primarily driven by three collaborations, each of which publishes its observed data in a variety of formats and repositories.
The H.E.S.S. Collaboration, observing in Namibia's Khomas Highlands, provides supplementary data to their published spectra.
The MAGIC Collaboration, observing in La Palma, Spain, provides access to a fits file repository containing data from observed spectra.
The VERITAS Collaboration, observing in Arizona, USA, recently established a dedicated online repository (VTSCat, see \cite{VTSCat}) to gather all of their published data in ECSV format. The latter is a format developed by the Astropy project\footnote{Astropy: Python library for astronomy, see \href{https://www.astropy.org/}{https://www.astropy.org/}} to store astronomical data as well as their associated metadata.

However, some publications are missing from those databases, and some spectra are reported with fluxes corrected for the EBL-induced absorption instead of the fluxes observed by the instruments.
Efforts have been made to create a unified repository of extragalactic VHE spectra, like BW15 and GammaCat, but these repositories are missing the most recent observations. 
To tackle this issue, we created the Spectral TeV Extragalactic Catalog, or STeVECat.
STeVECat compiles high-level results of VHE observations from 1992 to 2021, in the form of observed spectra and their associated observational metadata.

\section{STeVECat data}

\subsection{Data collection}

STeVECat is a database of extragalactic VHE spectra.
To compile the data in STeVECat, we searched journal publications referencing extragalactic sources in the TeVCat\footnote{TeVCat: online catalog for TeV astronomy; see \href{http://tevcat2.uchicago.edu/}{http://tevcat2.uchicago.edu/}} database.
We selected all publications providing TeV energy spectra with at least two spectral points.
This amounts to 173 publications, as opposed to 72 in BW15.
In some of those publications, the authors reanalyze previously published data; in such cases, we only report the most recent analysis in STeVECat.

Each individual spectrum is stored in a single ECSV file, and the files are ordered according to their associated publication.
The spectral data is saved as a table-like structure, with columns corresponding to the energy bin center and the associated flux and flux error.
We retrieved the spectral points for each selected publication from the publishing collaboration's repository when possible; otherwise, we contacted the relevant authors or digitized the figures. 
Additional columns appear when applicable, corresponding to the upper and lower edges of the energy bins, the negative and positive errors on the flux, and the upper limits on the flux.
The metadata, which is saved as a YAML-like structure, contains physical units associated with each column as well as the name of the observing instrument, the observation period in terms of start and end modified Julian date (MJD) as well as common calendar dates, the name and id of the observed source, the reference to the publication and a selection tag (see section \ref{sec:reliability}).
The livetimes, excess counts over background, and significance also figure in the metadata when they were provided in the original publication.

\begin{figure}[t]
    \centering
    \includegraphics[trim=0cm 1cm 0cm 0cm, width=0.8\linewidth]{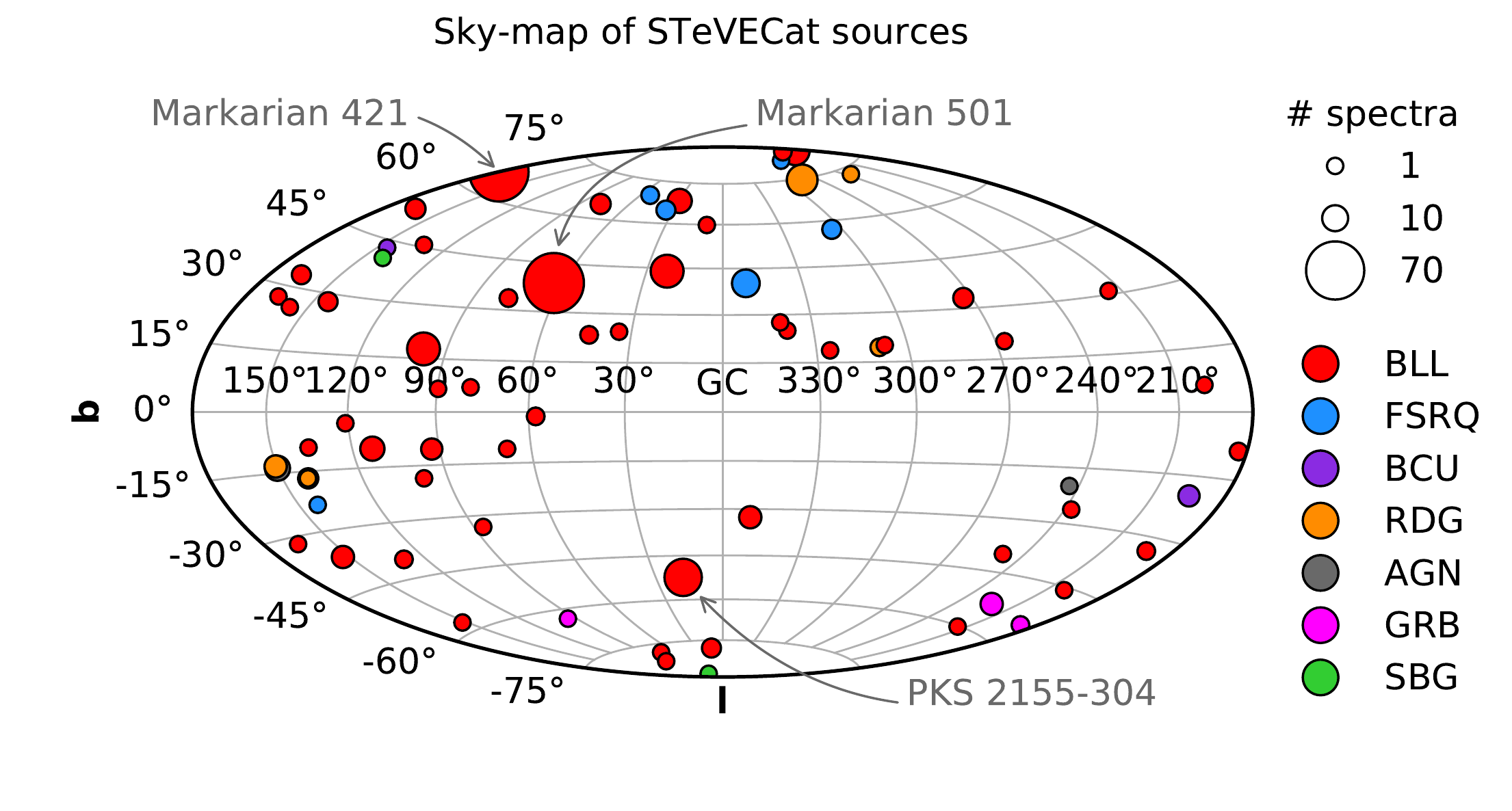}
    \caption{Sky-map of all sources referenced in STeVECat in Galactic coordinates. The area is a linear function of the number of associated spectra, and the color represents the source type: BL Lac (BLL), Flat spectrum radio quasar (FSRQ), blazar candidate of uncertain class (BCU), radio galaxy (RDG), AGN of unknown class (AGN), gamma ray-burst (GRB) or starburst galaxy (SBG). The three most observed sources are highlighted: Markarian 421, Markarian 501, and PKS 2155-304.}
    \label{fig:sky-map}
\end{figure}

STeVECat gathers 403 spectra from 73 extragalactic sources which are shown in Fig.~\ref{fig:sky-map} from publications between January 1992 and December 2021.
The majority of the sources are active galactic nuclei (AGNs) of the BL Lac category (BLL, see \cite{2022Galax..10...35P}), but the catalog includes as well three gamma-ray bursts (GRBs) observed in the TeV range. 

\subsection{Reliability of STeVECat}
\label{sec:reliability}

We performed checks on the extracted data to ensure the reliability of STeVECat.
We compared STeVECat data to data from GammaCat and VTSCat to ensure that STeVECat was complete for spectra observed by H.E.S.S., MAGIC and VERITAS. 
To ensure the readability of the files and the correctness of the physical units, we plotted each spectrum using GammaPy~\cite{GammaPy}, the Science Analysis Tool chosen by the Cherenkov Telescope Array Observatory, and compared it to the corresponding published figure.
In those comparisons, we paid close attention to the EBL-absorption of each spectrum.
When the spectra reported by collaborations were deabsorbed for the EBL effect, we included back the absorption using the same EBL model and source redshift as the authors.
All the spectra in STeVECat are hence observed spectra.
To facilitate contemporaneous analysis of the data, we ensured the validity of the reported observation period, which is the time in MJD of the first and last observation.

Some observations span long periods of time, several years in some cases.
The authors of such publications frequently report different sets of spectra, one set corresponding to the time-averaged spectrum over the entire observation period and another set corresponding to spectra associated with each sub-period of observation.
These spectra are not independent of one another, but we decided to include these two types of spectra in STeVECat.
To enable studies requiring independent spectra, we considered all the sets of spectra that provided the greatest temporal coverage, and annotated the one with the largest number of spectra.
In the following, we refer to this latter set of spectra as the non-overlapping set.
The non-overlapping set gathers 350 spectra from 153 publications.

\section{Using STeVECat}

\subsection{Information on the sources}

\begin{figure}[t]
    \centering
    \includegraphics[width=.75\linewidth]{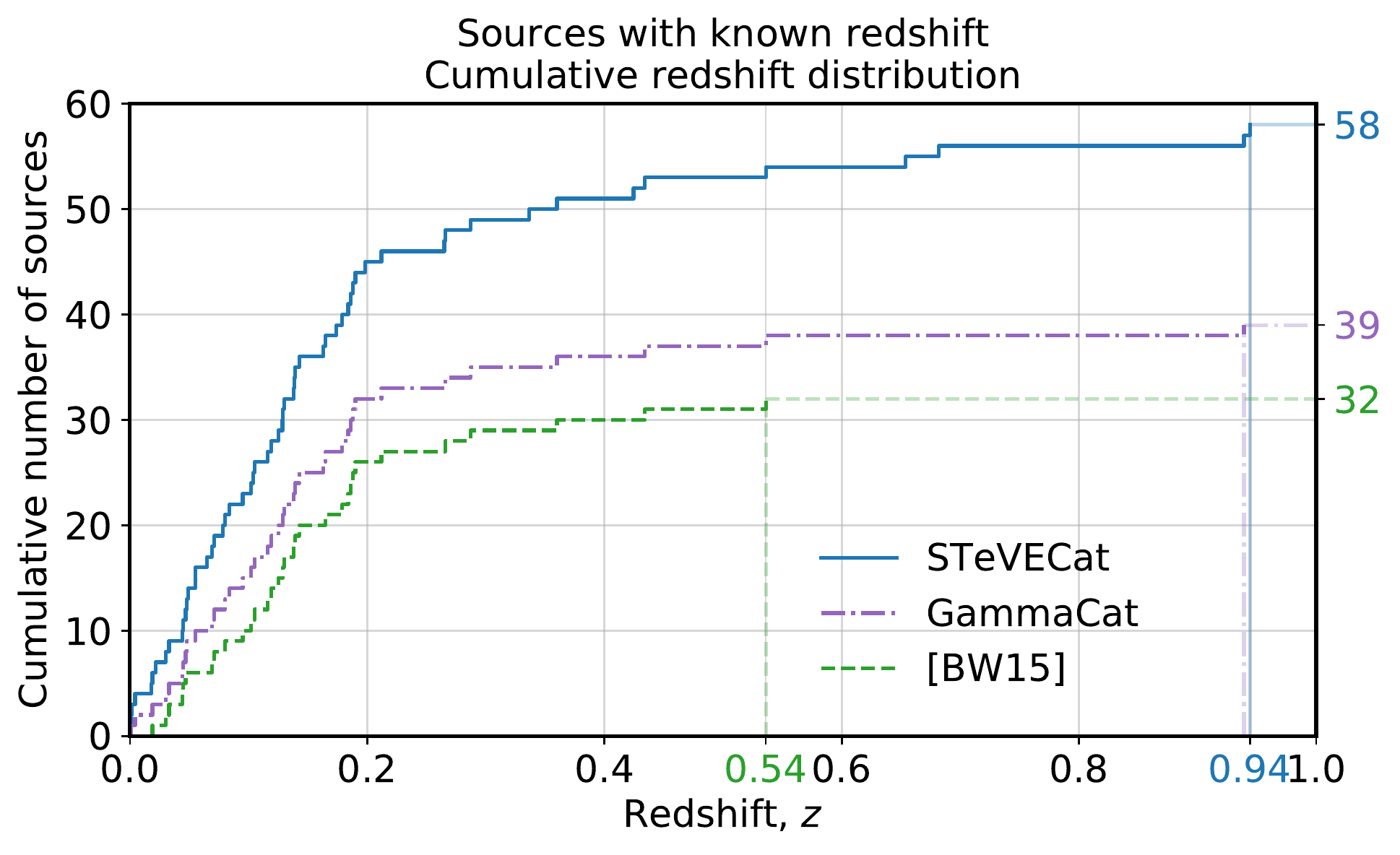}
    \caption{Cumulative redshift distribution for sources with known redshift. The blue line corresponds to sources referenced in STeVECat, the purple line corresponds to sources referenced in GammaCat, and the green line corresponds to sources referenced in the catalog from BW15.}
    \label{fig:redshift}
\end{figure}

To ease the use of STeVECat, we provide a table with information on all spectra and associated extragalactic sources in the catalog.
Following the conventions adopted by GammaCat and VTSCat, we report the assigned id that we use in the catalog for each source.
We also provide sky coordinates in right ascension and declination for each source using the SIMBAD Astronomical Database.\footnote{SIMBAD: astronomical database of objects beyond the Solar System, see \href{https://simbad.cds.unistra.fr/simbad/}{https://simbad.cds.unistra.fr/simbad/}}

When possible, we assign a redshift measurement to each source in STeVECat, either from a literature review or from dedicated spectroscopic redshift measurements \cite{G21}.
Each redshift has a tag corresponding to the reliability of the measurement (solid measurement, uncertain measurement, lower limit or unknown redshift).
STeVECat lists 58 sources with a solid redshift measurement.
In contrast, 39 sources with known redshift are represented in the catalog GammaCat,\footnote{GammaCat: source catalog for VHE $\gamma$-ray astronomy, see \href{https://gamma-cat.readthedocs.io/index.html}{https://gamma-cat.readthedocs.io/index.html}} whereas 32 sources with known redshift are included in BW15.
The cumulative redshift distribution for sources with known redshift from STeVECat, GammaCat and BW15 is shown in Fig.~\ref{fig:redshift}.

\subsection{Perspectives for EBL studies}

As explained in \cite{Gamma-ray_cosmo}, EBL studies using VHE gamma-rays require a large ensemble of extragalactic observations.
The previous study that used the largest VHE sample was BW15, with 90 spectra from sources with known redshift.
STeVECat collects 356 spectra from sources with known redshift in total, 310 spectra being in the non-overlapping set.
In comparison to previous studies, the size of the spectral sample for EBL studies has therefore more than doubled.

\begin{figure}[t]
    \centering
    \includegraphics[width=.75\linewidth]{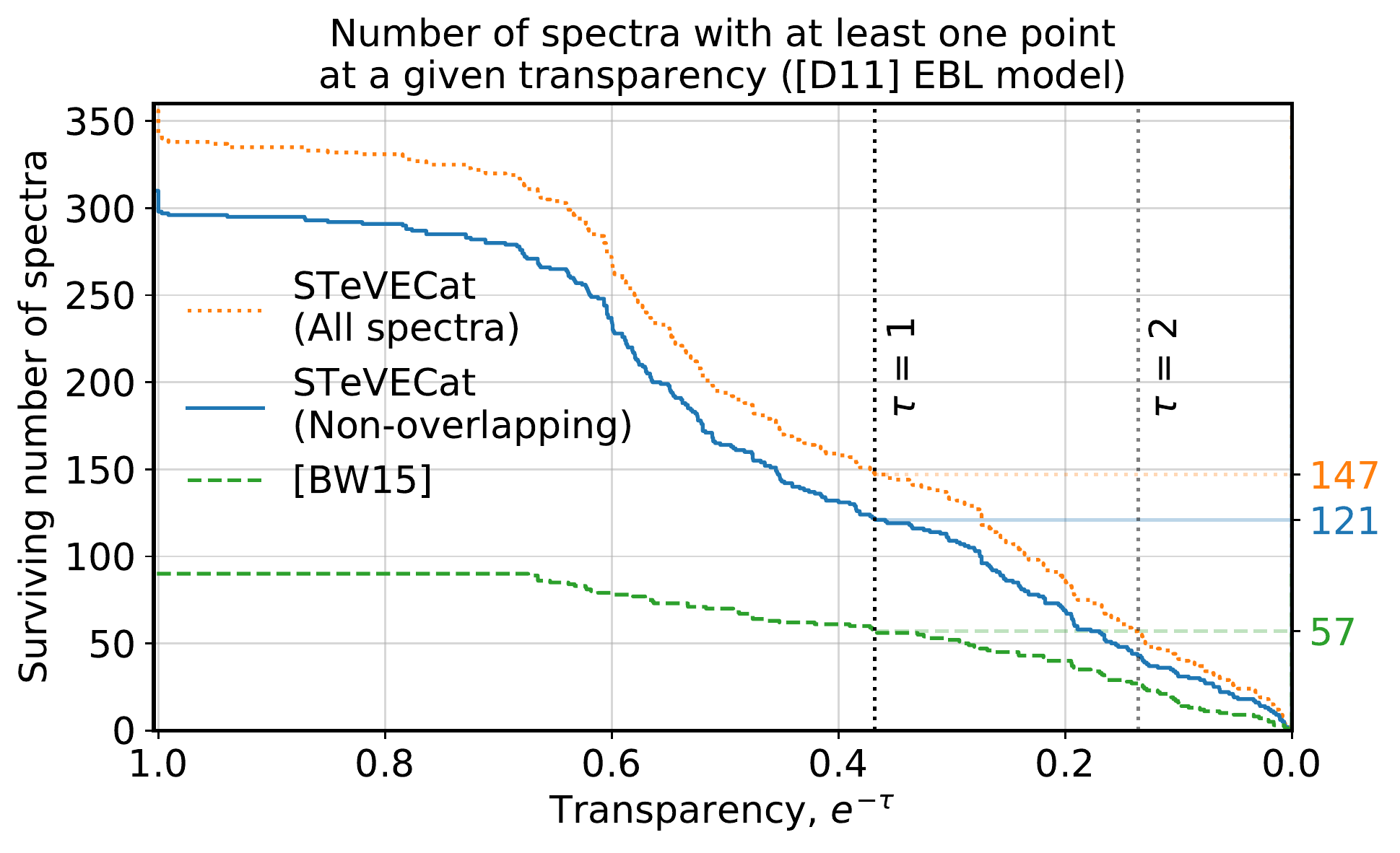}
    \caption{Surviving number of spectra from sources with known redshift reaching a given transparency for the EBL model from D11. The orange line corresponds to all the spectra from STeVECat, the blue lines corresponds to the non-overlapping spectra from STeVECat, and the green line corresponds to the spectra from BW15.}
    \label{fig:energy-reach}
\end{figure}

To highlight the prospects on EBL studies using STeVECat, we show in Fig.~\ref{fig:energy-reach} the surviving number of spectra from sources with known redshift, as a function of the transparency from the EBL model of \cite{D11} (hereafter D11).
For each transparency level, $e^{-\tau}$,  this corresponds to the number of spectra having a spectral point at this transparency level or below.
Using the EBL model of D11, 121 spectra from the non-overlapping set reach an optical depth $\tau = 1$, which is more than twice as many spectra as in BW15.
More information on the perspectives on EBL studies using the STeVECat corpus can be found in \cite{Pipeline}.

\section{Conclusion}

Since the first TeV detection of Markarian 421 in 1992, the VHE gamma-ray sky has grown significantly.
To enable analysis of the entire extragalactic TeV sky, we created STeVECat, the largest database of extragalactic VHE spectra to date.
STeVECat collects published extragalactic very-high energy spectra and associated observational meta-data.
The data is formatted in accordance with the standards used in publicly accessible repositories, and the entire catalog can be quickly loaded with GammaPy, the Science Analysis Tool chosen by the Cherenkov Telescope Array Observatory.

STeVECat collects 403 spectra from 173 journal publications, making it the most comprehensive collection of VHE extragalactic spectra to date.
Each of these spectra reports the observed spectrum, which means it can be used to infer EBL properties.
We provide a subset of spectra called the non-overlapping set, which corresponds to observations that can be considered independent, to enable population studies of extragalactic gamma-ray sources, studies of the GeV-TeV connection, and studies of absorption on the EBL.
STeVECat will be made publicly available.

\bibliographystyle{JHEP}
\bibliography{arxiv}

\providecommand{\href}[2]{#2}\begingroup\raggedright\begin{thebibliography}{10}

\bibitem{Punch_92}
M.~Punch, C.~Akerlof, M.~Cawley, M.~Chantell, D.~Fegan, S.~Fennell et~al.,
  \emph{{Detection of TeV photons from the active galaxy Markarian 421}},
  \href{https://doi.org/10.1038/358477a0}{\emph{Nature} {\bfseries 358} (1992)
  }.

\bibitem{2022Galax..10...35P}
E.~{Prandini} and G.~{Ghisellini}, \emph{{The Blazar Sequence and Its Physical
  Understanding}},
  \href{https://doi.org/10.3390/galaxies10010035}{\emph{Galaxies} {\bfseries
  10} (2022) 35}.

\bibitem{Gamma-ray_cosmo}
J.~{Biteau} and M.~{Meyer}, \emph{{Gamma-Ray Cosmology and Tests of Fundamental
  Physics}}, \href{https://doi.org/10.3390/galaxies10020039}{\emph{Galaxies}
  {\bfseries 10} (2022) 39}.

\bibitem{BW15}
J.~{Biteau} and D.A.~{Williams}, \emph{{The Extragalactic Background Light, the
  Hubble Constant, and Anomalies: Conclusions from 20 Years of TeV Gamma-ray
  Observations}}, \href{https://doi.org/10.1088/0004-637X/812/1/60}{\emph{\apj}
  {\bfseries 812} (2015) 60}.

\bibitem{NievasRosillo:2021aS}
M.~Nievas~Rosillo and T.~Hassan, \emph{{A data-driven evaluation of Fermi-LAT
  extrapolation schemes to the VHE regime.}},
  \href{https://doi.org/10.22323/1.395.0722}{\emph{PoS} {\bfseries ICRC2021}
  (2021) 722}.

\bibitem{VTSCat}
W.~{Benbow}, A.~{Brill}, M.~{Capasso}, J.L.~{Christiansen}, A.J.~{Chromey},
  M.K.~{Daniel} et~al., \emph{{VTSCat: The VERITAS Catalog of Gamma-Ray
  Observations}},
  \href{https://doi.org/10.3847/2515-5172/acb147}{\emph{Research Notes of the
  American Astronomical Society} {\bfseries 7} (2023) 6}.

\bibitem{GammaPy}
C.~Deil, J.~Lefaucheur, R.~Zanin, C.~Boisson, B.~Khelifi, R.~Terrier et~al.,
  \emph{{Gammapy - A prototype for the CTA science tools}},
  \href{https://doi.org/10.22323/1.301.0766}{\emph{PoS} {\bfseries ICRC2017}
  (2017) 766}.

\bibitem{G21}
P.~Goldoni, \emph{{Review of redshift values of bright AGNs with hard spectra
  in 4LAC catalog}},  Apr., 2021.
\newblock 10.5281/zenodo.5512660.

\bibitem{D11}
A.~{Dom{\'\i}nguez}, J.R.~{Primack}, D.J.~{Rosario}, F.~{Prada},
  R.C.~{Gilmore}, S.M.~{Faber} et~al., \emph{{Extragalactic background light
  inferred from AEGIS galaxy-SED-type fractions}},
  \href{https://doi.org/10.1111/j.1365-2966.2010.17631.x}{\emph{\mnras}
  {\bfseries 410} (2011) 2556}.

\bibitem{Pipeline}
L.~{Gréaux} and J.~{Biteau}, \emph{{TeV Bayesian Study of the Extragalactic
  Background Light}},  in \emph{These proceedings}.

\end{thebibliography}\endgroup

\end{document}